# Preprint: Bigdata Oriented Multimedia Mobile Health Applications

Zhihan Lv*, Javier Chirivella, Pablo Gagliardo

**Abstract** This is the preprint version of our paper on JOMS. In this paper, two mHealth applications are introduced, which can be employed as the terminals of bigdata based health service to collect information for electronic medical records (EMRs). The first one is a hybrid system for improving the user experience in the hyperbaric oxygen chamber by 3D stereoscopic virtual reality glasses and immersive perception. Several HMDs have been tested and compared. The second application is a voice interactive serious game as a likely solution for providing assistive rehabilitation tool for therapists. The recorder of the voice of patients could be analysed to evaluate the long-time rehabilitation results and further to predict the rehabilitation process.

## 1 Introduction

Big data in healthcare and mobile health (mHealth) refers to electronic health data so large and complex that they are difficult to manage with traditional software, computing methods or data management tools. Health data in the background of big data includes clinical data (medical imaging, physicians advices and notes, pharmacy, insurance), electronic patient records (EPRs) or electronic medical records (EMRs), health monitoring data (heart beat, blood pressure, daily walking distance and activities) and social media data (Twitter feeds, Facebook post [4]) [63]. Regarding to clinical data, researchers categorized it into clinical records, health research records and business/ organization operations records [21]. The objective of applying big data in healthcare and mHealth is achieving higher quality care at lower costs and in better overall outcomes. By analyzing patient characteristics and the cost and outcomes of healthcare to identify the most cost effective treatments and hospital, thereby influencing

———————————————
Z. Lv*, J. Chirivella, P. Gagliardo
FIVAN, Valencia, Spain
*Corresponding Author. E-mail: lvzhihan@gmail.com

provider behavior, applying advanced analytics to patient, broad scale disease profiling to identify predictive events and support prevention initiatives, collecting and publishing data on medical procedures, assisting patients in determining the care protocols or regimens that offer the best value, predicting and minimizing fraud by implementing advanced analytic systems, licensing data to assist pharmaceutical companies in identifying patients for inclusion in clinical trials, application of big data in mHealth can improve the quality of health care quality. Mobile Health (mHealth) is fast becoming an exciting new prospect in the world of medicine, especially combined with big data, mHeath shows particular potential benefits for low and middle income countries. The World Health Organization (WHO) first described and analyzed the concept of mHealth in a report published in 2011 [33]. The first academic studies focused on mHealth were conducted within the past six years. Several papers have presented the results of pilot projects on: aid or emergency lines [10] [29] [64], appointment reminders by SMS [19], reminders to take anti-malarial or antiretroviral medication [60], support to stop smoking or improve physical activity [64], prevention of risk behaviors among youth, follow-up of diabetic or asthmatic patients [20], or mobile telemedicine and mobile phone support in collecting health data [6], use mobile technology for behavior change communication (BCC) [26].

Successful application of mHealth with big data, such as researchers implemented a remote health consultancy service over a "portable clinic" and a software tool, "GramHealth" [1] and a study on the novel cloud supported health awareness system to detect the abnormal data in the large scale Wireless Body Area Networks (WBANs) using MapReduce infrastructure [61], encourages many more studies on different fields in medicine. Studies conducted in developed countries have reported individuals being more receptive and positive towards mHealth enhancing patient self-management of long term diseases by therapy adjustments, supportive messages, and medication and appointment reminders [27]. Prognotive Computing recognize patterns and formulates its own structure to provide a solution or gives a predicted alert has been studied, the system provides a handle of Health care and life span of numerous life forms, apart from recognized human health patterns with adaptive algorithms it is able to predict the endangering of patients [70].

In mHealth, performance and availability of data processing are critical factors that need to be evaluated since conventional data processing mechanisms may not provide adequate support, a study proposed a generic functional architecture with Apache Hadoop framework and Mahout for handling, storing and analyzing big data that can be used in different scenarios [14]. Problems of personal data security on mHealth already tackled by numerous researchers. A study proposes a geographical privacy-access continuum framework, which could guide data custodians in the efficient dissemination of data while retaining the confidentiality of the patients/individuals concerned [18]. Chronical disease management is a prospers field of mHealth application. Hypertension and Diabetes are rapidly growing chronic conditions in Pakistan with an increasing mortality rate due to poor knowledge, self-management and low quality of life; the potential positive and high scope of using Mobile Health services for management in chronic conditions like Diabetes and Hypertension in a developing country is tested with mobile phones, the results showed that Mobile phones can help bridge healthcare gaps, but facilitating safe informal m-health for young people should be a policy priority [28].

Data mining on the clinical data with the help of mHealth also accomplished great results. Big Data drawing on Medicare claims and IRIS Registry records can help identify additional areas for quality improvement in cataract surgery and care in ophthalmology [8]. Intelligent handling of the cost effective mHealth data can accomplish the goals of one health to detect disease trends, outbreaks, pathogens and causes of emergence in human and animals [2]. A study developed a new mobile-based approach to automatically detect seizures, the approach is experimentally evaluated using the standard clinical database [52]. This paper introduce two mHealth applications, which can be employed as the terminals of bigdata based health service to collect information for electronic medical records (EMRs).

## 2 Application

In this section, we introduce two multimedia mobile health applications, which are oriented to mobile client of health bigdata. The two applications respectively interacted with users by 3D perception and voice.

### 2.1 Immersive VR Glassess

The first application is a novel immersive entertainment system for the users of hyperbaric oxygen therapy chamber. The system is a hybrid of hardware and software, the scheme is described in this paper. The hardware is combined by a HMD (i.e. virtual reality glasses shell), a smartphone and a waterproof bag. The software is able to transfer the stereoscopic images of the 3D game to the screen of the smartphone synchronously. The comparison and selection of the hardware are discussed according to the practical running scene of the clinical hyperbaric oxygen treatment [40]. Finally, a preliminary guideline for designing this kind of system is raised accordingly.

The efficacy of hyperbaric oxygen therapy has been recorded in a lot of literature in clinical research community [71] [59] [16] [62]. The limited space of the hyperbaric oxygen chamber has constrained the possibilities of activities. The most common activity that the users usually do in the hyperbaric oxygen chamber is sleeping, although the oxygen generation machine of the chamber is noisy which couldn't be ignored at all. Playing with smartphone or reading book also happens sometime, since they only need activities of upper arms. It's however still not comfortable to retain the postures of the upper arms and neck to play or read for long time. The physical discomforts may impair the psychological enjoyment. Overall, the spatial limitation constrains the physical activities inside the hyperbaric oxygen chamber. Our research plans to improve the user experience of hyperbaric oxygen therapy by novel virtual reality (VR) based interactive technologies, the imagination of the running scenario is shown as in figure 1, where the patient lies in the hyperbaric oxygen chamber and wear the virtual reality immersive glasses.

Virtual reality (VR) environments are increasingly being used by neuroscientists [5] and psychotherapists [66] to simulate natural events and social interactions.

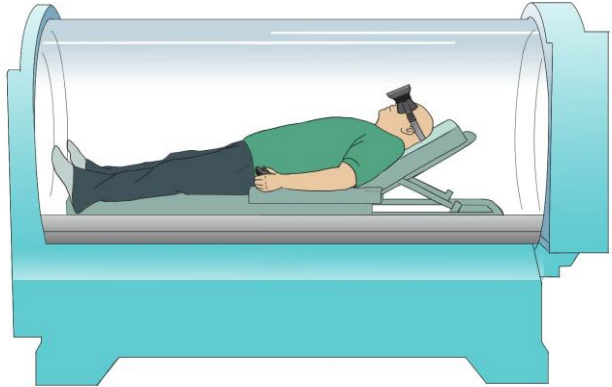

**Fig. 1** The system running scenarios.

VR technology has been proved to be able to stimuli for patients who has difficulty in imagining scenes and/ or are too phobic to experience real situations since long time ago [56]. The 'Immersive' characteristic of VR technology can substantially improve movement training for neurorehabilitation [54] [9] [34] [3]. The report of the early research of utilizing VR to treat claustrophobia patient [7] has inspired our work, since the hyperbaric oxygen chamber is a typical sealed space. The underwater VR game for aquatic rehabilitation brings us some suggestions and tips [32] about the development of VR game on head mounted device (HMD) in unusual air environment (e.g. under water, in hyperbaric oxygen). Nonetheless, VR in clinical rehabilitation is still in infancy and need more exploration [57]. In addition, the medical sensors may enhance the related research in future [81] [37] [79].

*2.1.1 Hardware*

The VR hardware has experienced several-generation development and emerged a number of useful immersive environment, which ranges from early surround-screen projection-based VR environment so-called CAVE [13] [12] to latest human-computer-interface (HCI) such as VR glasses. In our case, CAVE is impossible to be taken into the hyperbaric oxygen chamber due to the high cost and the dependence on plane context. Therefore, VR glasses become the best choice for bringing the 3D immersive perception to users lying inside hyperbaric oxygen chamber.

The size of the hyperbaric oxygen chamber is 185CM length, 90CM diameter, in which an adult cannot sit up. The chamber is built by plastic sheeting for insulation and has four small windows on each side, as shown in figure 3 (C). Since the electronic device is not allowed to directly use in the hyperbaric oxygen chamber in order to prevent the explosion, the smartphone has to be sealed into a insulation bag which could separate the phone to the high-density oxygen. In our designed system, we employ a waterproof bag to wrap the smartphone and then put it into the HMD, as shown in figure 2. The smartphone has Quad-Core 2.5GHz CPU, 3GB DDR3 memory, the screen is 1920*1080 resolution, 441PPI, 95% NTSC color gamut, it supports Wifi and bluetooth. The HMD shells we have chosen include four productions as shown in Figure 4.

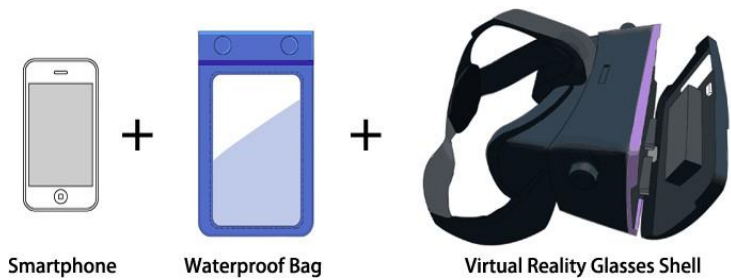

**Fig. 2** Waterproof bag is employed to separate the phone to the high-density oxygen.

### 2.1.2 Software

The smartphone selected in our system runs Android operating system. Other smartphone operating systems are also available in future development.

Two kinds of software technologies can implement the stereoscopic 3D applications on smartphone screen, as described as followed. Virtual Desktop. As shown in Figure 3, the 3D game GZ3DOOM which is the modified version of DOOM is running in stereoscopic 3D mode and the game window is duplicated to the smartphone screen by a screen-synchronization software. The connection between the PC and smartphone is by Wifi. In our system, we employed Trinus VR [76] as the screen-synchronization software, which includes a .exe program running on windows as server side and a .apk application on smartphone as client side. Some other similar software can be accessed from google play too, such as Intugame VR [24]. It worth to mention that, TrinusVR has fake3D function which can generate fake stereoscopic image by automatically duplicating the original image for each eye. Even though the fake3D actually brings flat perception for user, there are also tools that make the conversion from monoscopic of 3D game to stereoscopic [39] side by side (SBS), like Vireio [55] or TriDef 3D [72]. Mobile Phone APP. The 3D game or VR scene application running on smartphone may suffer worse system performance. The available applications resources are not sufficient too since the accumulated 3D games on PC or other game devices in past cannot directly run on smartphone or be simply modified to be suitable for smartphone operating system (i.e. Android). Anyway, the customized mobile APP developed for smartphone has better compatibility with the hardware and probably will be improved by next-generation Cloud operating system such as Windows 10, which is a promising development trend of operating system. The VR glasses SDK (i.e. Cardboard) for latest version of video game engines (i.e. Unity3D) have been provided to adapt existing Unity3D APPs for vir-tual reality [25] [50], which is a available choice with friendly user interface for non-programmer. The wireless communication may be improved by novel network technologies [30] [80] [31] [15].

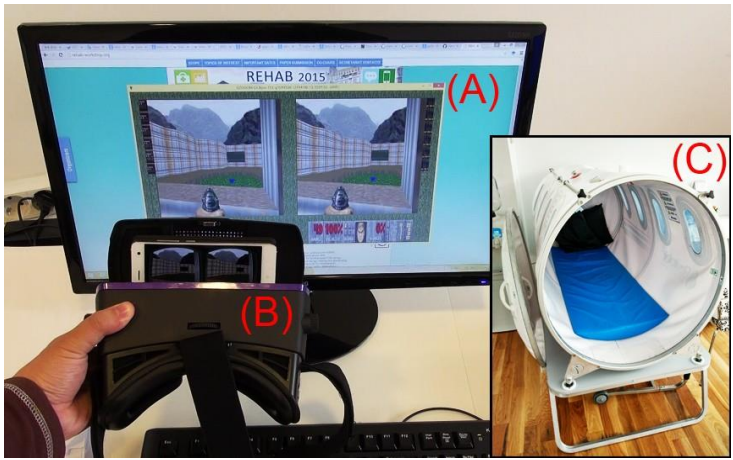

**Fig. 3** The graphical content of screen (A) is duplicated to smartphone screen (B). (C) is the hyperbaric oxygen chamber located in our clinic.

### 2.1.3 Running Scenarios

The user (e.g. patient, athlete, healthy people) is lying inside the hyperbaric oxygen chamber, wears the HMD and holds the remote controller as shown in figure 1. The user can play the 3D game in stereoscopic 3D mode or 2D game in fake3D mode which can also show the game image on the VR glasses. Moreover, conventional need of reading E-books or accessing Internet are still possible to be achieved on VR glasses by fake3D mode. The input methods include head motion and remote controller. The head motion on VR glasses only supports the rotations around three axis which are controlled by gyroscope sensor of the smartphone. The head rotation actions are synchronous with the rotation of the avatar's view in the VR scene [51], which brings the immersive perception to the user in real time. Meanwhile, the remote controller is used to input the displacement of the avatar as well as manipulate the menu of the game configuration. The gamepad based manipulation may be replaced by touch-less interaction [46] [47] [45] [44].

### 2.1.4 Preliminary Comparison

Figure 4 shows the HMD that we have considered and compared for the clinical need. Where (a) [53](c)(d) are the VR glasses shell by which users could watch the anaglyph 3D scene generated from smartphone screen. While (b) is so-called 'Lazy Glasses', the lens of which reflects the scene toward the below of user's eyes. As shown in figure 5, the user can lie flat and play the smartphone instead of raising the head or forearms. As we have known, (b) is not a VR technology based device, but it has better suitability. The user could read book or play smartphone game by wearing while lifting the neck or raising arms are needless any more (b). Moreover, as a non-electronic device, (b) is more safe and convenient in hyperbaric oxygen chamber. However, (b) couldn't generate the stereoscopic 3D images for immersive perception which is the essential condition of VR. Anyway, (b) is a convenient choice for users who don't really desire to enjoy 3D immersive experience while still hope to read in the hyperbaric oxygen chamber.

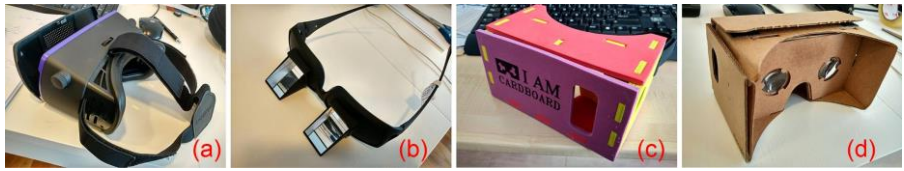

**Fig. 4** The HMD that we have considered and compared for the clinical need.

Comparing (a)(c)(d), the three shells are respectively made by hard plastic, ethylene-vinyl acetate (EVA), and cardboard. The convergence-to-face part of (a) for light blocking is made by soft holster filling of sponge, so it's not oppressive at all. In addition, (a) can modify the pupil distance (PD) and depth of field (DOF), so (a) is suitable for the users with different myopia degree and PD. (c) and (d) have not adequate adaptive functions for users with disparity in physical characteristics, both are however in collapsible structures the portability of which leads to convenience for distribution and transportation. In our practical clinical need, the features of (b) are more suitable for different patients. By the way, the price level of the four devices we have chosen are almost equal, so the cost problems will not affect the motivation of the device choice in future practical application.

The common drawback of the HMDs is the dizziness. Dizziness is mainly caused by the vision delay which leads to vestibular and cochlear imbalance. In our case, lying posture doesn't lead to more dizziness than other postures such as standing and sitting. So we believe the VR glasses will bring immersive enjoyment to hyperbaric oxygen chamber users as long as it can create pleasure to common players. The image enhancement tehchnologies would improve the vision feedback effect of the software part [82] [77] [78].

*2.1.5 Designing Guideline*

1. The water proof bag is necessary for isolation from hyperbaric oxygen.
2. It's significant to choose a comfortable HMD shell.
3. The smartphone programming doesn't have to be done as long as a PC or laptop is nearby as a server.
4. An additional remote controller is needed for menu manipulation.
5. Dizziness is not enhanced by hyperbaric oxygen chamber.
6. 'Lazy Glasses' is a non-electrical choice for users who hope to read 2D content.

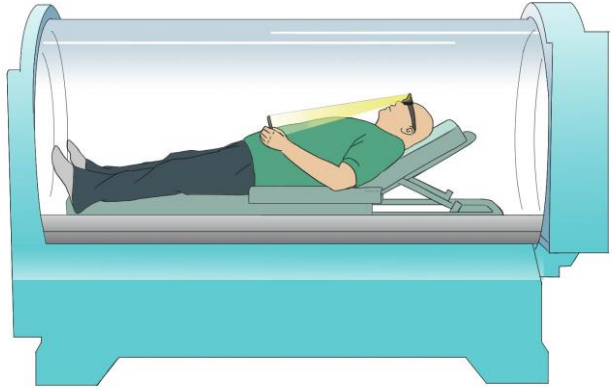

**Fig. 5** The user is experiencing 'Lazy Glasses' and playing with mobile phone.

2.2 Voice Training Serious Game

The second application is an assistive training tool for rehabilitation of dysphonia/presbyphonia, which is designed, developed and iteratively evaluated and optimized according to the practical clinical needs. The assistive tool employs serious games as the attractive logic part, and running on the tablet with normal microphone and/or Kinect2 as voice input device. The patient is able to play the game as well as conduct the voice training simultaneously guided by therapists at clinic, while not be interfered, or do rehabilitation exercise independently at home. The voice information can be recorded and extracted for evaluation of the long-time rehabilitation progress. This paper outlines a design science approach for the development of an useful software prototype of such a tool. Both therapists and sufferers have provided earnest suggestions for the improvement. Our application software detects the pitch and/or loudness as the evaluation factor. Several pitch estimation algorithms have been evaluated and compared with selected patients voice database, the best one selected according to the benchmark. The technical details are recorded in our previous publications [43] [41] [42].

In clinical, dysphonia is measured using a variety of tools that allow the clinician to see the pattern of vibration of the vocal folds, principally laryngeal videostroboscopy, such as coustic examination and electroglottography. Subjective measurement of the severity of dysphonia is carried out by trained clinical staff (e.g. GRBAS (Grade, Roughness, Breathiness, Asthenia, Strain) scale, Oates Perceptual Profile). Objective measurement of the severity of dysphonia typically requires signal processing algorithms applied to acoustic or electroglottographic recordings [38]. Besides invasive or drug based treatments, effective logopedic treatments have been proved [68]. However, constant training is a key factor for this type of therapy [35]. The clear speech recognition is not expected. The patients usually have inabilities of pronunciation of the high pitches. The hopeful assistive tool is able to retrieve the quantified long-time voice rehabilitation information.

### 2.2.1 System User Interface

According to the therapists' suggestions and our clinical observation, considering the design factors simultaneously, we designed an serious game based assistive tool with a user interface, the first version of the full mode (Therapist Mode) is as in figure 6 left. In this mode, the game logic (top-left), voice analysis (top-right), patient case history (bottom-left) and the game level editor (bottom-right) are embedded in the main window. The patient case history and the level editor are implemented by database management user control, it's rapid-developed and modifiable which is suitable for our gradual research and iterative development. The voice analysis UI visualizes pitch (red), loudness (purple) by dense histogram, and voice source by a semicircular compass. In this version, through setting the voice source direction on the semicircular compass, only the voice spreaded from the patient location could be received by the system, while all of noised or other interference voice including therapist voice will be filtered.

The software is able to estimate the basic information of patient voice (i.e. pitch (Mel) and/or loudness (dB)). In the first version, the frequency is identified by FFT (Fast Fourier Transform) based frequency-domain approach which turns the voice wave into a frequency distribution. Further more, the pitch (tone) and midi number are calculated depend on frequency.

As suggested by the therapist, pitch (Mel) is the main controlable factor, thus we uses it as the manipulating tool which function is equal to the joystick in the classic video game. Meanwhile, the game UI renders the vision feedback of the user manipulation. Some other factors are also considered, such as Phonation time (ms), pitch change (Mel), Duration (s), Reaction time (ms).

Built based on the considered factors, the configuration includes some essential parameters which can modify the difficult level of the games. The parameters contain sensitivity, x spread, y spread, incoming speed, voice maintenance, session duration. The parameters are available to modify by the level editor, and the new level will be generated for the specific patients as soon as it's saved. The designed functions can provide immediate evaluation feedback to therapists and patients. One space flight game is created as the logic part of the assistive tool, as shown in figure 6. In this game, the player uses voice pitch to control the spaceship as the avatar to dodge the planets. The spaceship moves up if the pitch is higher than 200, and move down if lower than it, the value is adjustable. Once the spaceship collides the planets, the game is over and the player score is recorded. Considering the manipulation complexity for the dysphonic patients, the loudness factor isn't utilized in the game interaction.

Beside the designed game stories, some assisting functions are implemented, include 'create patient', 'record and replay voice', 'save therapy'. In our gamification therapy, one game session is one therapy, that means when the therapist starts a game session for patient, the therapy starts synchronously.

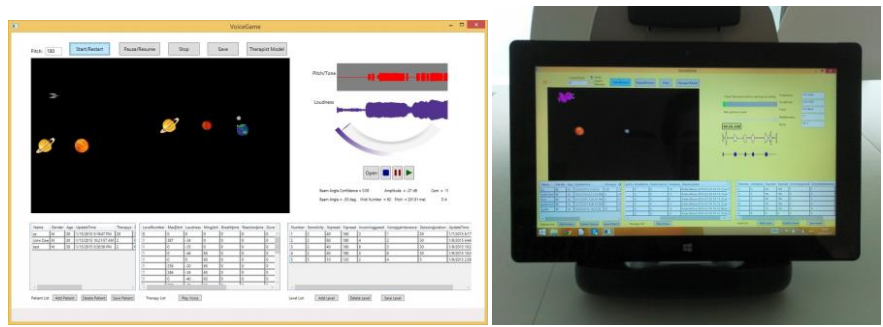

**Fig. 6** Left: The full model of the system running scene; Right: New user interface of the proposed tool which is suitable for tablet with normal microphone.

*2.2.2 Running Scene*

In the first version, when the software is running, both therapist and patient stand in front of Kinect2 and watch the monitor as in figure 7. The kinect2 can recognize the patient by location. During the game, the patient will control the game avatar to follow the game role, at the same time, the therapist makes voice to supervise the patient. The implied gamification feature during this process is that the patient try the best to follow the therapist's voice guide. When the game is over, the software will save the biometric data automatically.

The second version of the software supports common microphone and gets rid of Kinect2. Because we have realized that the treatments are like the driving training, the independent driving isn't expected at the early training usually. Especially when the game is out of control but the patient cannot increase the pitch at all, the therapist should rapidly make voice and help the patient to complete the game, which gives patient a paradigm and can reduce anxiety and boost confidence of the patients. The new user interface is shown as in figure 6 right. The left side of the UI is the game logic, the right side is the estimated pitch values monitor which presents frequency, pitch, amplitude, midi note, midi note number, sample, duration.

## 3 Conclusion

The recent multiplication of mHealth projects worldwide illustrates the overall trend towards the globalization and technologization of biomedicine, but problems still remain unsolved in the following aspects. The crucial part in healthcare and mHealth big data is EMRs. Though report argues that healthcare sector has been much slower than other industries, such as finance and retail, in transitioning to and utilizing computerized information systems, like EMRs [21]. For example, in 2009, 99% of primary care physicians in the Netherlands used EMRs. In contrast, 36% of primary care physicians in Canada and 46% of primary care physicians in the United States used EMRs [22]. Until 2013, only 78 percent of office-based physicians used some form of electronic medical record system [23]. Some large health insurance providers and systems, such as the Veterans Health Administration, Kaiser Permanente, and Cambridge Health Alliance, offer personal health records that allow patients to access their electronic record for appointments, prescription requests, and other services [65]. With expected dramatically growth of healthcare data in the heading years [11], the slow pace of informatization and integrate degrees formed major obstacles in the application of big data in healthcare and mHealth. Besides the slow utilizing, the inter-organizational operability of EMRs is also less common within a country or region [17]. And concerns about personal data security and anonymity still hinders the application massive application of mHealth [58]. The new contributions in cloud computing, bigdata and network will bring more new production to mHealth field [83] [75] [86] [36] [73] [74] [69]. The connection with braincomputer interface (BCI) application is also a promising topic in future [84] [85].

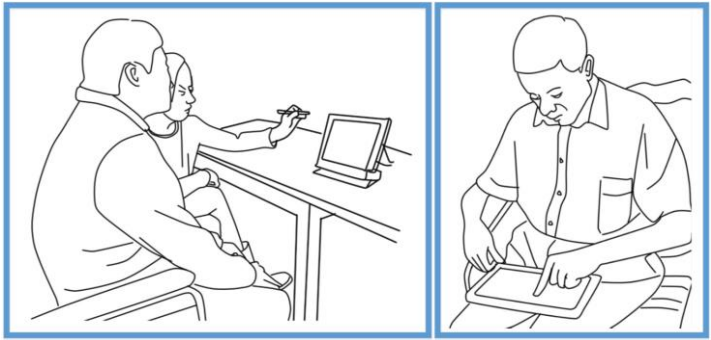

**Fig. 7** Running scene.

In this paper, we proposed two mHealth applications which can collect patient information as electronic medical records (EMRs). The first one is a hybrid system for improving the user experience in the hyperbaric oxygen chamber. By this system, users can enjoy 3D stereoscopic games wearing virtual reality glasses and immersive perception, as well as read E-books or play the 2D games. Several HMDs have been tested and compared by us. All the technical issues have been solved, the new challenges are the customized 3D games or software used for this case. The efficacy of this system will be evaluated in future research by subjective measurement such as Geneva Emotion Wheel (GEW) [67] as well as the medical scales [7] (e.g. Subjective units of discomfort scale (SUDS), Problem-related impairment questionnaire (PRIQ), Self-efficacy towards the target behaviour measure (SETBM), The attitude towards CTS measure (TAM)). Moreover, the technology will be employed accompany with the balance measurement [48] [49]. The contributions of the second application is developing a voice interactive serious game as a likely solution for providing assistive rehabilitation tool for therapists. The recorder of the voice of patients could be analysed to evaluate the long-time rehabilitation results, as well as the prediction of the rehabilitation process. In the next stage, we will consider the efficacy of the proposed system. Finally we will test the safety of the system.

**Acknowledgment**

The authors would like to thank Sonia Blasco, Vicente Penades and Chantal Esteve for their fruitful help and suggestions. The work is supported by LanPercept, a Marie Curie Initial Training Network funded through the 7th EU Framework Programme under grant agreement no 316748.